\documentclass{article}

\usepackage[final]{neurips_ai4d3_2023}
\usepackage{multirow}

\usepackage[utf8]{inputenc} 
\usepackage[T1]{fontenc}    
\usepackage{hyperref}       
\usepackage{url}            
\usepackage{booktabs}       
\usepackage{amsfonts}       
\usepackage{amsmath}
\usepackage{nicefrac}       
\usepackage{microtype}      
\usepackage{xcolor}         
\usepackage{graphicx} 
\usepackage{placeins}
\usepackage{svg}

\title{Assessing interaction recovery of predicted protein-ligand poses}

\author{
David Errington \quad Constantin Schneider \quad C\'edric Bouysset \quad Fr\'ed\'eric A. Dreyer$^{*}$\\
Exscientia, Oxford Science Park, Oxford, OX4 4GE, UK \\
\texttt{\{derrington,cschneider,cbouysset\}@exscientia.co.uk} \\
\texttt{dreyer.frederic@gene.com}
}

\begin{document}

\maketitle
\def\thefootnote{*}\footnotetext{Currently at Prescient Design, Genentech, New York City, USA.}

\begin{abstract}
The field of protein-ligand pose prediction has seen significant advances in recent years, with machine learning-based methods now being commonly used in lieu of classical docking methods or even to predict all-atom protein-ligand complex structures. Most contemporary studies focus on the accuracy and physical plausibility of ligand placement to determine pose quality, often neglecting a direct assessment of the interactions observed with the protein. In this work, we demonstrate that ignoring protein-ligand interaction fingerprints can lead to overestimation of model performance, most notably in recent protein-ligand cofolding models which often fail to recapitulate key interactions.

\end{abstract}

\section{Introduction}

Recent advances in AI-based docking hold the potential to generate accurate protein-ligand poses at often a fraction of the computational cost of classical docking algorithms. Additionally, cofolding models that can directly predict the full protein-ligand complex structure have emerged as a promising alternative, circumventing the need for docking while providing the capability to model conformational changes to the protein.

As these machine learning (ML) methods are typically trained on the PDBBind General dataset~\citep{pdbbind} released in 2020, it has become commonplace to benchmark them using the PoseBusters test suite~\citep{posebusters} which consists of 308 protein-ligand complexes released after 2021 and that are, therefore, outside their training data. 

It has previously been noted~\citep{cole_comparing_docking,posebusters,posecheck,baillif2024} that ML methods lack the necessary inductive bias to generate realistic poses, even though they can often obtain low root-mean-squared deviation (RMSD) values from the crystal structure ground truth. Performing further quality checks on the ligand chemistry and the physical plausibility of the pose, notably through the PoseBusters benchmark, is therefore an important test for ML-based docking tools.

However, from the perspective of computational chemists, a physically plausible pose with low RMSD is a necessary but not sufficient condition for that ligand to be of interest. In particular, these conditions ensure that the ligand is close to where it should be and adopts a sensible pose within the pocket, but for that pose to be of biological relevance, it must also create key interactions between the protein and the ligand. These interactions are in fact often used to constrain classical docking tools, an option that is not currently available in ML docking methods. 
Such interactions are typically classified using protein-ligand interaction fingerprints (PLIFs), which identify the protein residue, the interaction type and, optionally, the ligand atom involved in the interaction. Several tools exist to detect PLIFs~\citep{plip,arpeggio,oddt,ichem} and in this work we use the ProLIF package~\citep{prolif}.

\begin{figure}
    \centering
    \includegraphics[width=0.52\linewidth]{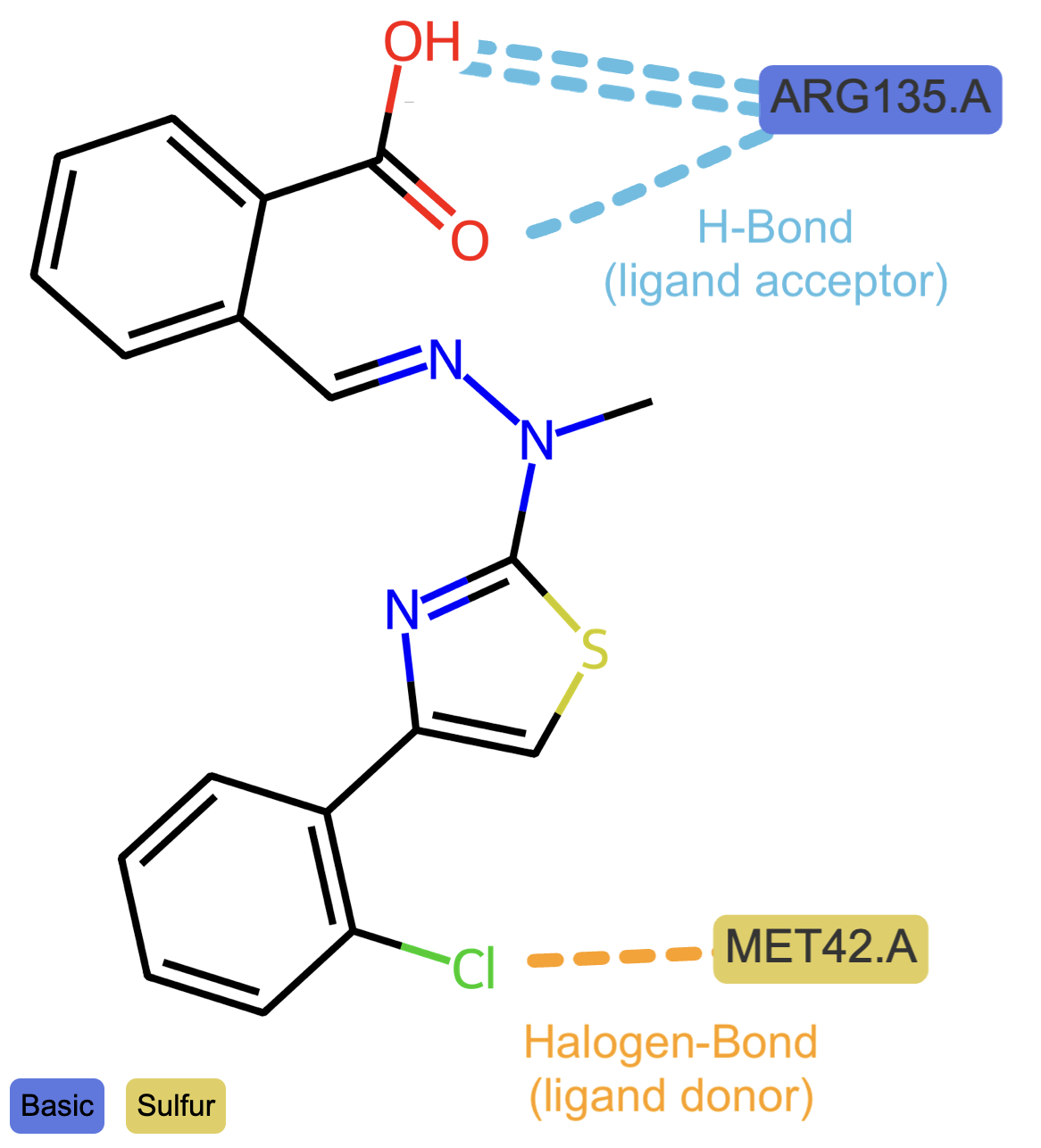}\qquad
    \includegraphics[width=0.42\linewidth]{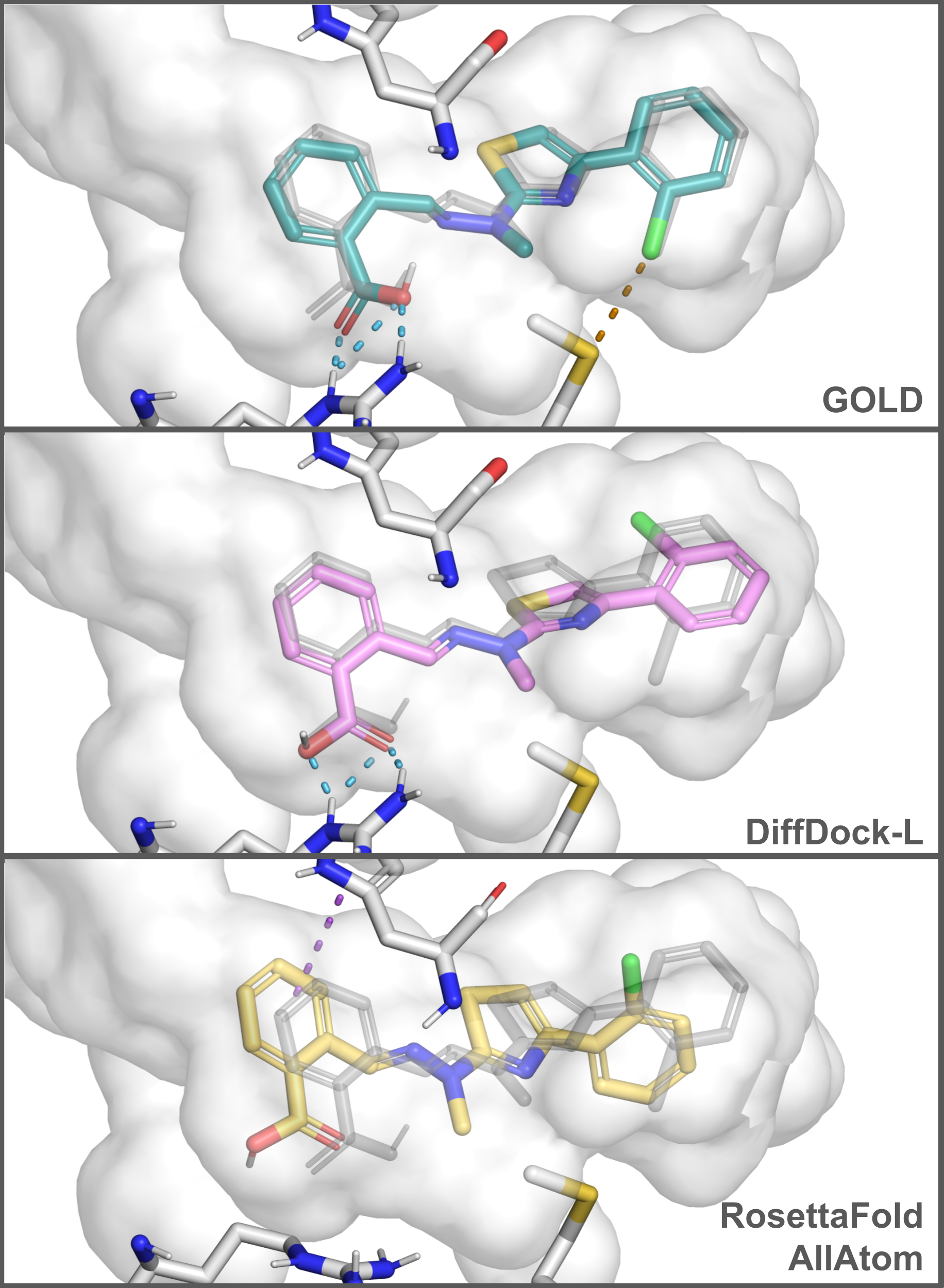}
    \caption{Left: Two-dimensional representation of the ligand EZO and its four interactions with the crystal structure 6M2B. Basic residues are shown in blue and residues containing a sulfur atom are shown in yellow. Right: Docked poses generated with GOLD, DiffDock-L and RosettaFold-AllAtom showing the calculated interactions for each model, with the ground truth ligand in grey.}
    \label{fig:plif_example}
\end{figure}

In Figure \ref{fig:plif_example}, we show on the left a visualisation of the PLIFs detected in the crystal structure of the protein target 6M2B with ligand EZO, and on the right the 3D poses generated by GOLD (classical docking), DiffDock-L (ML docking) and RoseTTAFold-AllAtom (ML cofolding). The ground truth complex has hydrogen bonds and a halogen bond. In this example, both GOLD and DiffDock-L are able to identify PoseBuster-valid (PB-valid) poses with RMSD$\leq$2Å, but whilst DiffDock-L recovers 75\% of the PLIFs from the crystal pose, missing the halogen bond interaction with the Chlorine atom, GOLD is able to recover all of them. DiffDock-L also changes the conformation of the ligand so that the hydrogen bonding involves a different set of atoms, while GOLD recovers the exact ground truth pose. RoseTTAFold-AllAtom meanwhile, which has the more challenging task of also reconstructing the protein, finds a pose with a RMSD of 2.19Å and steric clashes, which also fails to recover any of the ground truth crystal interactions. 

ML methods do learn indirectly about protein-ligand interactions but without an explicit term to this effect in the loss function, the training  signal is weak, and ML docked ligands can often end up with key functional groups pointing in the wrong direction. 
In contrast, classic docking algorithms are, through the design of their scoring functions, inherently interaction-seeking; their top scoring poses are those that achieve certain key interactions. In this paper, we aim to motivate PLIF recovery as a useful metric for assessing model quality and use them to benchmark a number of modern pose prediction tools.

\section{Method}

\subsection{Protein-ligand interaction fingerprint}\label{sec:prolif}

Interaction fingerprints summarize the three-dimensional interactions present in a molecular complex.
In the context of small molecule drug discovery, we are primarily interested in interactions that a ligand achieves with the protein pocket of interest, for which PLIFs provide a vectorized representation.
This representation typically consists of a mapping between protein residues and a ligand along with a bitvector that can encode different types of interactions, such as hydrophobic, $\pi$-stacking, $\pi$-cation, ionic, and hydrogen bonds.
PLIFs were calculated with the ProLIF package~\citep{prolif} considering only hydrogen and halogen bonds (donor and acceptor), $\pi$-stacking, cation-$\pi$ and $\pi$-cation, and ionic interactions (anionic and cationic), excluding the less specific hydrophobic interactions and Van der Waals contacts. These  are more rarely considered by computational chemists as key interactions that must be recapitulated as they are non-directional and therefore highly correlated with RMSD. This is because these latter interactions are much more promiscuous than the others, which would result in a weaker signal from polar interactions despite their critical importance in ligand-protein binding. Custom distance thresholds were used for hydrogen bonds (3.7Å), cation-$\pi$ (5.5Å) and ionic (5Å) interactions while all other parameters are the defaults in ProLIF~v2.0.3.

We note that while not all the other PLIF-calculation tools previously mentioned require explicit hydrogens to be present in the input files~\citep{plip,arpeggio}, they end up adding them if not present, although the optimisation of the hydrogen bond network is either not enabled by default, or not available.

The interactions detected by PLIF-libraries are very sensitive to the protonation state of both the protein and the ligand as it can decide whether an interaction gets labelled as ionic or hydrogen bond. Whilst the classical docking methods can model hydrogens explicitly, their scoring functions often infer the potential for interactions such as hydrogen bonds from the geometry of the heavy atoms alone. Meanwhile, ML methods typically model only heavy atoms. In order to treat all methods equally, we place explicit hydrogens on the protein structure using PDB2PQR~\citep{pdb2pqr}, as well as on the ligand pose if not already present, using RDKit~\citep{rdkit}. We then performed a short minimisation of the ligand inside the pocket, defined as protein residues within 6Å of the ligand, whilst keeping the heavy atoms fixed, using RDKit's implementation of the Merck Molecular Force Field (MMFF)~\citep{rdkit_mmff,mmff94s}. This is a consistent way to optimise the hydrogen bond network of the docked/cofolded pose and gives each method the best possible chance to make interactions from the proposed heavy atom positions.

\subsection{Classical docking algorithms}\label{sec:classical}

Classical molecular docking aims to predict plausible ligand poses when binding to a protein target, leveraging computational algorithms to accurately simulate molecular interactions, as pioneered by the development of the DOCK~\citep{seminal_docking_Kuntz1982} and AutoDock~\citep{autodock} algorithms. In this analysis, we use the FRED, HYBRID and GOLD algorithms which are more modern approaches to classical docking.

FRED and HYBRID are docking programs from the OEDocking suite~\citep{oedock} and are rarely included in ML docking benchmarks. Both algorithms work by first generating an ensemble of conformations which then undergo rigid docking into a specified pocket. FRED is an unbiased docking program that uses only the structure of the target protein to position and score molecules, whilst HYBRID is a biased docking program that also uses the structure of the reference ligand to find the optimal docked pose~\citep{fred_hybrid_benchmark}. HYBRID is typically used in a lead optimisation campaign to dock novel compounds that differ minimally from a reference ligand. For the self-docking task we consider in this work, HYBRID has an unfair advantage over the other methods and we include it here mainly to validate this advantage over FRED.

Finally we include CCDC GOLD~\citep{gold}. Unlike the OEDocking tools, GOLD generates ligand conformations on the fly as it places the ligand in the pocket. 

FRED and HYBRID both use the ChemGauss4 scoring function~\citep{oedock} whilst GOLD uses the PLP scoring function~\citep{plp} to identify the optimal pose. In both cases, these scoring functions pay close attention to the shape and hydrogen bond complementarity of poses within the active site. In contrast to ML methods, classical docking methods explicitly seek interactions and we hypothesise this will lead to improved PLIF recovery and ultimately more favourable poses. 

For all three classical methods we return 10 poses and then select the pose with the top docking score for our subsequent analysis.

Additionally, we note that existing benchmarks in the literature often perform classical docking with minimal processing of the PDB files, overlooking refinement steps to address issues like missing loops, alternate conformations, flipped functional groups, and adding explicit hydrogen atoms to the ligand and protein structures consistently with their titration states. A suitable preparation of input files ensures that the active site residues and the ligand are ready for docking, making the simulations more accurate and predictive of ground-truth interactions. Since we are using OpenEye docking tools in this work, we performed structure preparation using the Spruce CLI from OpenEye ~\citep{oespruce}. We note however, that other structure preparation tools do exist such as Reduce~\citep{reduce}, the CSD API from CCDC~\citep{csd_api_ccdc} and the Protein Preparation Wizard from Schr\"{o}dinger~\citep{schrodinger2024}.

\subsection{ML docking algorithms}
The application of ML to accelerate molecular docking and find more accurate binding poses has received a lot of interest in recent years~\citep{equibind,tankbind,unimol}.

In this work, we consider DiffDock-L~\citep{diffdock}, the latest version of DiffDock which uses confidence bootstrapping to improve significantly on previous versions.  DiffDock-L is a state-of-the-art ML docking model that uses a diffusion model over the non-Euclidean manifold parameterizing the ligand degrees of freedom in order to generate plausible orientations and conformations. DiffDock-L uses its confidence model to assign a score to each sampled pose and so, as with the classical methods, we sample 10 poses for each ligand and use the highest-confidence pose for our subsequent analysis. 

It is worth highlighting that the confidence model underpinning DiffDock-L is a GNN classifier trained to identify poses with RMSD$\leq$2Å and, whilst this will indirectly capture some information about interactions, it does not explicitly rank poses based on PLIFs in the same way as classical scorers.

\subsection{Protein-ligand cofolding}
Several structure prediction models have recently incorporated the description of more general biomolecular assemblies beyond simple protein polypeptide chains, including the capability of cofolding a protein and small molecule simultaneously and predict all corresponding atomic coordinates~\citep{rfaa,umol,af3,neuralplexer}. 
We test two such models, Umol~\citep{umol} and RoseTTAFold All-Atom (RFAA)~\citep{rfaa}. Unlike the docking methods described in previous sections, Umol and RFAA will return a protein different to that in the crystal structure and the protein structure is also an output of the model. 

Cofolding is a complex problem and, whilst there has been much progress recently, it is still a relatively nascent field. As a result, it is not uncommon for the output structures to have issues such as steric clashes, overabundance of cis-peptide bonds or gaps in the protein, or to fail to preserve the chemistry of the input ligand (\textit{e.g.}, flipped stereochemistry). By default, Umol performs postprocessing in a attempt to fix this whereby it generates conformers of the input ligand and then returns the conformer with the best Kabsch alignment against the predicted atom positions. This approach guarantees the chemistry of the output ligand matches the chemistry of the input ligand. Finally, Umol then places hydrogens and uses OpenMM~\citep{openmm} to optimise the protein-ligand system and it is this optimised complex that we use in our subsequent analysis. 

In contrast, RFAA does no such postprocessing out of the box. Consequently, we often find that the output ligand either has invalid stereochemistry or it has valid stereochemistry (making it PB-valid) but this stereochemistry is different to that of the input ligand. To ensure we are assessing the correct ligand, and for consistency with Umol, we add a similar postprocessing pipeline to RFAA, but with minimization in the YASARA2 forcefield~\citep{yasara}, which we found to be more tolerant than OpenMM to unphysical structures.

\subsection{Data and Metrics}

The original PoseBusters test suite identified 308 high-quality protein-ligand complexes released after 2021 and therefore outside the training data of most ML methods~\citep{posebusters}. We excluded 37 data points due to limitations in compute time for cofolding involving large targets, and 8 due to either structure preparation or forcefield failures. A further 7 targets were found to have no relevant interactions in the crystal pose, which arises when the ligand and pocket residues exclusively have hydrophobic interactions which we do not calculate, or the interactions in the complex are slightly outside of the distance and angles thresholds used to generate PLIFs. Altogether, this leaves 256 PoseBuster complexes for our analysis.

We run each of the methods on the PoseBusters dataset and record the following properties
\begin{itemize}
    \item RMSD to crystal pose
    \item PoseBuster validity
    \item PLIFs of the predicted pose
\end{itemize}
The central contribution of this paper is the introduction of a PLIF recovery rate metric. This metric measures the percentage of interactions in the crystal pose that are successfully replicated in the docked or cofolded pose, as measured by PLIFs generated by ProLIF (see Section \ref{sec:prolif}), and captures how well each method can account for protein-ligand interactions.

\section{Results}

We now turn to the evaluation of interaction recovery in predicted ligand poses with classical docking, ML docking and protein-ligand cofolding on the PoseBusters dataset.

\subsection{PoseBusters benchmark}

\begin{figure}
    \centering
    \includegraphics[width=0.7\linewidth]{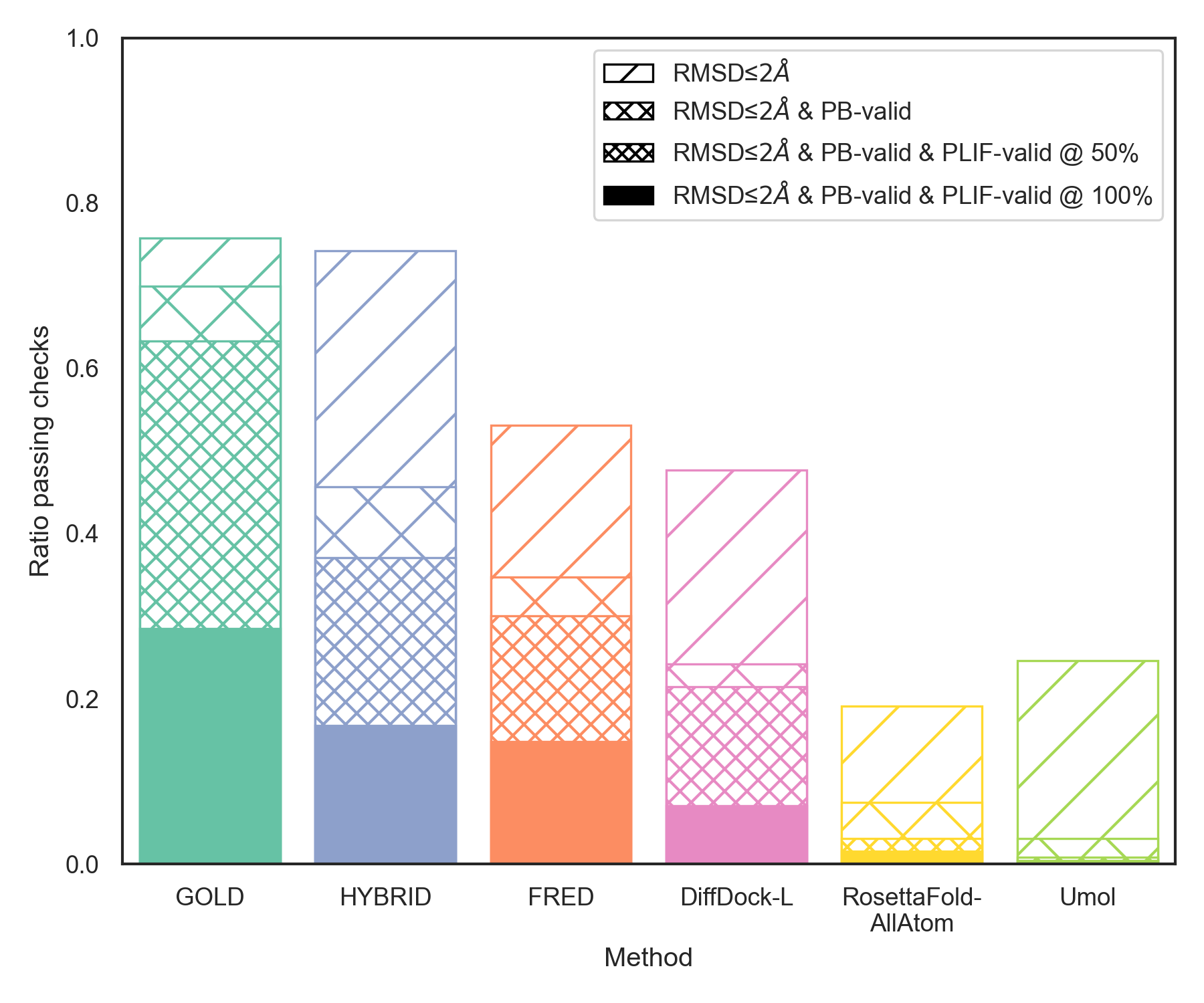}
    \caption{The ratio of predicted protein-ligand complex structures for each model passing checks on ligand positioning (RMSD$\leq$2Å), physicality (PoseBuster-valid) and interaction recovery (PLIF-valid). }
    \label{fig:passing_checks}
\end{figure}

Figure~\ref{fig:passing_checks} shows the overall results of the six methods on the PoseBusters benchmark set. As in the original PoseBusters paper~\citep{posebusters}, we show performance according to different metrics. The striped region shows the percentage of poses with RMSD$\leq$2Å, whilst the coarse crosschecked region shows the percentage of poses that are also physically plausible and successfully pass the PoseBuster validity checks. Our new additions are the fine crosschecked and solid regions which show the percentage of poses that are additionally ``PLIF-valid'' and succeed in also recovering at least 50\%  and 100\% of the interactions present in the crystal pose respectively.

One observation in the original PoseBusters study is that by RMSD alone, the original DiffDock model is shown to outperform GOLD but, when physical plausability metrics are added, GOLD is substantially better. However, Figure~\ref{fig:passing_checks} shows that in our study, GOLD does substantially better than the latest DiffDock-L model, even on the RMSD criteria. This is because we perform structure preparation on the protein before docking as described in Section~\ref{sec:classical}. This is more typical of how traditional docking tools are used in a drug discovery campaign, while classical methods are often used somewhat naively when benchmarking ML algorithms~\citep{inductivebio_blogpost_af3_100linesofcode}. 

Turning our attention to the full set of results, it is clear that the three traditional docking algorithms outperform the three ML algorithms across every metric, with GOLD achieving the best results. Indeed, GOLD finds more poses successfully recovering at least 50\% of the crystal interactions than any of the ML methods are able to produce falling within $2$Å RMSD. As expected, HYBRID outperforms FRED due to its ability to use prior knowledge from the crystal ligand pose. Interestingly though, despite being the only method to have this prior advantage, HYBRID is still outperformed by GOLD.

We find that cofolding methods achieve substantially worse interaction recovery than DiffDock-L. Umol achieves a higher fraction of ligands placed within $2$Å RMSD than RFAA though it should be noted that, unlike RFAA, Umol receives pocket residues as input. However, Figure~\ref{fig:passing_checks} shows that for both cofolding tools, but especially for Umol, the vast majority of these poses are physically implausible and missing key interactions.

\subsection{Interaction recovery rates}

Whilst the previous section focused on the number of poses that successfully recovered either 50\% or 100\% of the PLIFs in the crystal pose, here we look at the distribution of PLIF recovery rates across all PoseBuster data points. 

\begin{figure}
    \centering
    \includegraphics[width=\linewidth]{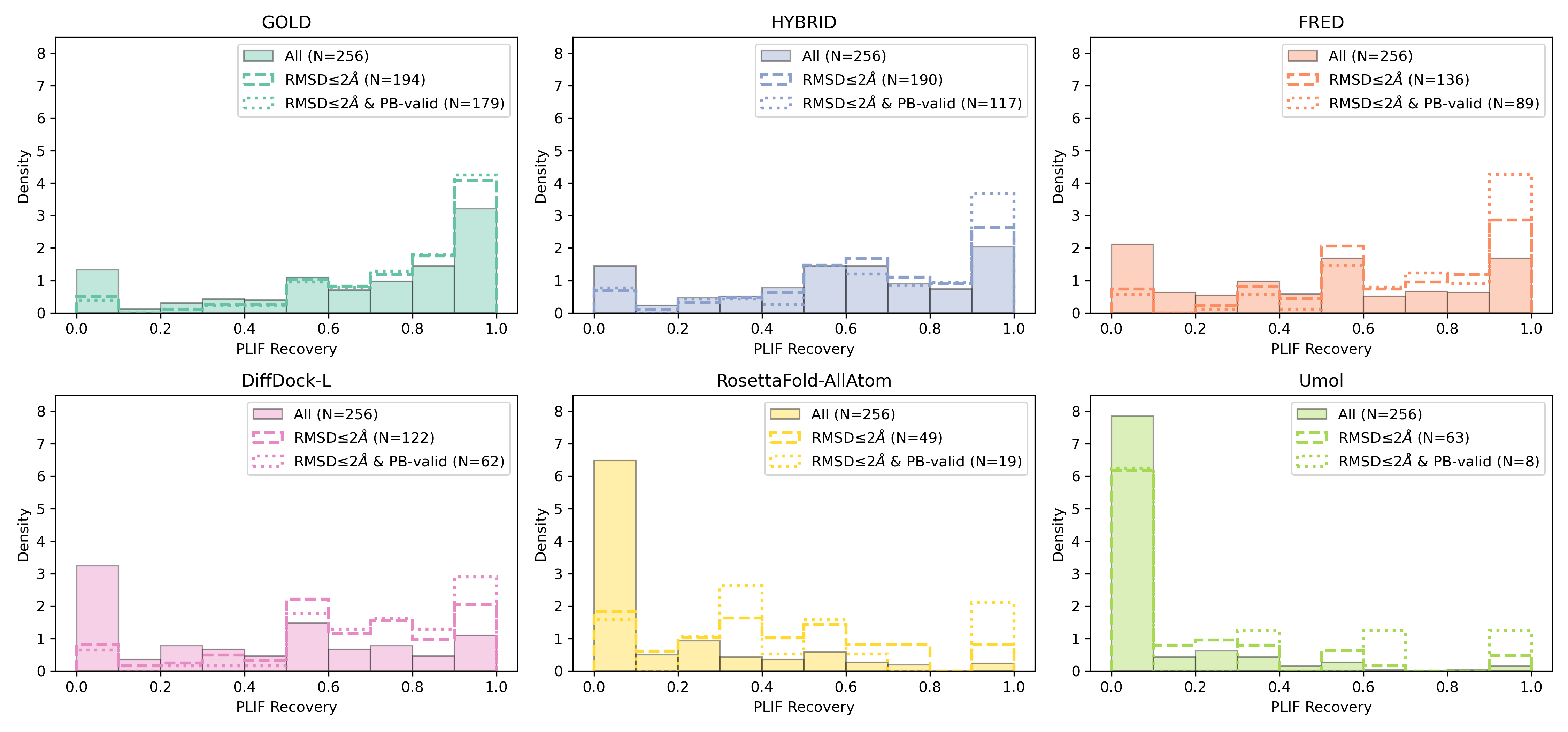}
    \caption{Recovery of protein-ligand interaction fingerprint for each model. The distribution of PLIF recovery among poses that pass the RMSD and PoseBuster test are shown in dashed and dotted lines.}
    \label{fig:recovery_rates}
\end{figure}

In Figure~\ref{fig:recovery_rates} we show a histogram of PLIF recovery rates for every method. We use normalized histograms to highlight the impact on this distribution of the RMSD and PoseBuster validity criteria.

We see a noticeable difference in skew between the histograms for the classical methods and the histograms for the ML methods, confirming that classical methods are much more successful at recovering the crystal interactions.

Under the premise that protein-ligand interactions are what we are actually interested in, we can ask the question whether either the RMSD$\leq$2Å filter or the RMSD$\leq$2Å and PB-valid filter are sufficient to leave only poses that make key interactions. If so, we would see a large change in the skew of the histogram as we apply these filters as poses with low PLIF recovery would get filtered out. We observe a noticeable change to all distributions when applying the RMSD filter, which removes ligands placed too far from the ground truth pose for any interactions to be recovered. In the case of GOLD, the PLIF recovery rate is relatively unaffected by the PB-valid filter. The change in skew is more noticeable in HYBRID, FRED and ML-based methods, though the latter have a sample size after filtering too small to be conclusive. It is however clear that many poses with few recovered PLIFs remain after these filters, confirming that interaction recovery can provide a useful orthogonal metric to PoseBuster validity.
Further analysis of the correlation between RMSD and PLIF recovery is shown in Appendix~\ref{app:scatter}.

\subsection{Recovery of different interaction types}

\begin{figure}
    \centering
    \includegraphics[width=\linewidth]{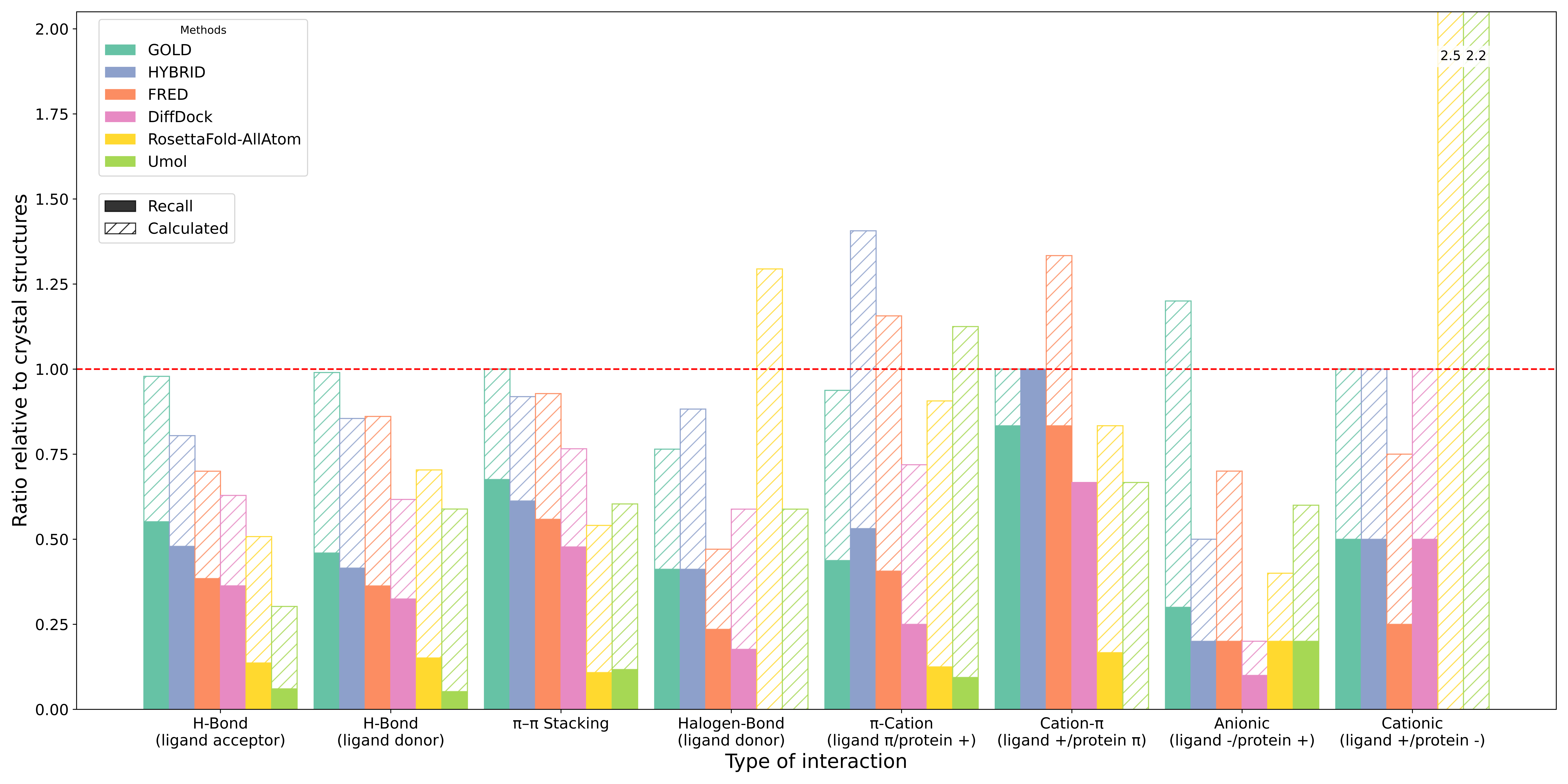}
    \caption{Ratio to the ground truth of calculated and correctly recovered (recall) interactions shown separately for each interaction types.}
    \label{fig:interaction_types}
\end{figure}

Up until this point in our analysis we have not distinguished between different types of protein-ligand interactions. In Figure~\ref{fig:interaction_types} we show a breakdown by model of the predictions for different types of interactions. The solid region shows the recall for each type of interaction whilst the striped region shows the ratio of detected PLIFs in the proposed pose relative to the PLIFs in the crystal pose.

Looking at the solid regions, we see that the classical methods produce poses that are better at recovering every type of interaction being considered with the exception of cationic interactions where DiffDock outperforms FRED. We hypothesise that this is because classical methods have scoring functions that explicitly seek interactions. 

Hydrogen bonds are the most important kind of interactions to consider~\citep{guide2interactions} and, as shown in Table~\ref{tab:xtal_plif_freq}, they are the most prevalent in our dataset, so it is worth emphasising the difference in recall observed across models in this case. It was previously noted that ligands produced by ML \textit{generative} methods do not make as many hydrogen bonds as found in reference datasets~\citep{posecheck}. Our results here confirm that this is also true for the simpler task of ML \textit{docking} where the reference ligand is given and the model is simply tasked with finding the optimal pose. Again, the reason that ML methods consistently recover fewer hydrogen bonds than classical methods is likely because the scoring functions driving classical methods are carefully optimised to prioritise hydrogen bonds. 

Turning to the striped bars, we can also observe that ML methods generally produce much fewer hydrogen bonds and $\pi$-stacking interactions, which are the most frequent interactions in ligand-protein docking as shown in Table~\ref{tab:xtal_plif_freq}.
The outliers in calculated cationic interactions for RosettaFold-AllAtom and Umol are due to a completely different orientation of the docking pose with respect to the crystal ligand, often replacing cation-$\pi$ interactions found in the crystal structure with cationic interactions.

\begin{table}[h]
    \centering
    \caption{Number of interaction types across PoseBuster crystal structures}
    \begin{tabular}{l@{\hspace{20pt}}r}
    \toprule
    Interaction & Frequency \\
    \midrule
    H-Bond (ligand acceptor) & 843 \\
    H-Bond (ligand donor) & 496 \\
    $\pi$-$\pi$ Stacking & 111 \\
    Halogen-Bond (ligand donor) & 17 \\
    $\pi$-Cation (ligand $\pi$ / protein $+$) & 32 \\
    Cation-$\pi$ (ligand $+$ / protein $\pi$)& 6 \\
    Anionic (ligand $-$ / protein $+$) & 10 \\
    Cationic (ligand $+$ / protein $-$) & 4 \\
    \bottomrule
    \end{tabular} \label{tab:xtal_plif_freq}
\end{table}

\section{Discussion}

In this paper, we have considered interaction fingerprints in protein bound small molecules. 
It has become commonplace to consider both ligand RMSD and PoseBuster validity as a proxy for model accuracy.
These metrics however do not fully capture the recapitulation of key interactions.
We studied how accurately different protein-ligand pose prediction tools, notably classical docking, ML docking and protein-ligand structure prediction models, can recover ground truth interactions.
PLIF recovery provides a useful metric, orthogonal to those used in existing benchmarks, which can further assess validity of predicted poses and is particularly valuable in drug discovery applications.

We showed that classical docking algorithms tend to substantially outperform ML-based methods in generating physically plausible poses, and recover relevant interactions with much higher success rate.
This result highlights the fact that classical docking benchmarks are rarely run competitively in the literature.
In contrast, cofolding models, where the coordinates of all atoms of the protein and ligand are jointly predicted, while often placing the ligand in the right location, rarely generate physically plausible poses that recover meaningful interactions with the target protein~\citep{cofolding_learn_interactions}.
Protein-ligand structure prediction is a harder task than docking, and also claims a much wider set of use cases, such as being able to adapt the conformation of the protein to accommodate different ligands or accurately model cryptic pockets, where the druggable pocket is absent in the apo structure and becomes exposed through interaction with the ligand~\citep{Oleinikovas2016,Meller2023a,Meller2023b}. However our results here suggest that in order for this emerging technique to be successful, considerably more attention is needed to ensure the predicted poses form key interactions. 
This could be achieved by incorporating an explicit PLIF or pharmacophore-sensitive loss to the training of ML models.
We note that it is possible to infer all interactions, including hydrogen bonds, from the geometry of the heavy atoms only and so we see potential to introduce geometric terms to the loss functions of ML methods to encourage this. Another simpler option would be to use a weighted RMSD that assigns a higher contribution to atoms matching specific pharmacophoric features (\textit{e.g.} hydrogen bond donors and acceptors, charged atoms, and $\pi$-rings).

The code used in this study is made available online at~\url{https://github.com/Exscientia/plif_validity}, along with all prepared protein structures at~\url{https://doi.org/10.5281/zenodo.13843798}.

\section*{Acknowledgements}
We are grateful to Henry Kenlay, Daniel Cutting, Gail Bartlett, Daniel Nissley, Luk\'a\v{s} Pravda, Ben Butt,  Richard Bradshaw, Francis Atkinson, Douglas Pires and Hagen Triendl for useful discussions.

\clearpage
\appendix
\section{Correlation betweeen PLIF recovery and RMSD}
\label{app:scatter}

\begin{figure}[ht]
    \centering
    \includegraphics[width=\linewidth]{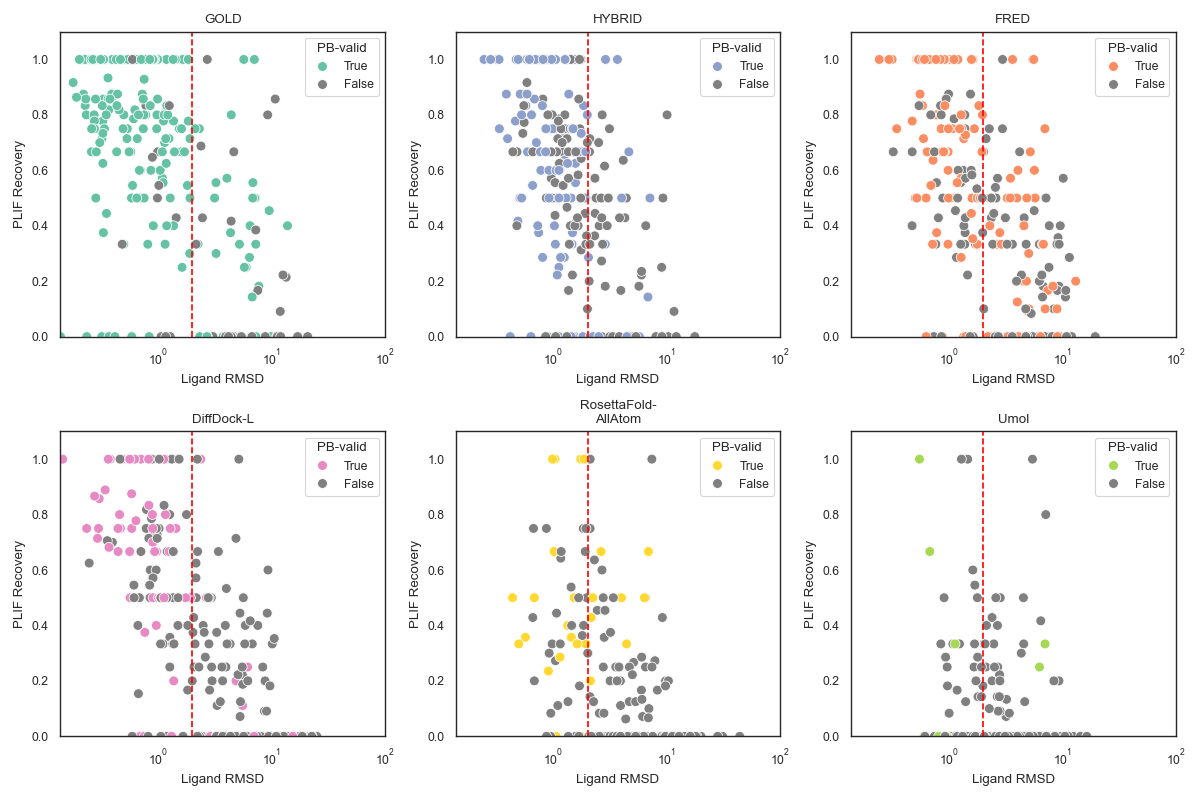}
    \caption{PLIF recovery rate and RMSD, highlighting data points which are PoseBuster-valid. Note that we use a modified definition of PB-validity that excludes ligand RMSD. The red line indicates a ligand RMSD of 2Å.}
    \label{fig:scatter}
\end{figure}

In Figure~\ref{fig:scatter}, we show a scatter plot of the PLIF recovery rate against the RMSD.

\clearpage

\bibliography{references}
\bibliographystyle{plainnat}


\end{document}